\documentclass[nohyperref]{article}

\usepackage{microtype}
\usepackage{graphicx}
\usepackage{subfigure}
\usepackage{booktabs} 

\usepackage{hyperref}
\usepackage{aas_macros}

\usepackage[accepted]{icml2022}

\usepackage{amsmath}
\usepackage{amssymb}
\usepackage{mathtools}
\usepackage{amsthm}

\usepackage[capitalize,noabbrev]{cleveref}

\theoremstyle{plain}

\theoremstyle{definition}

\theoremstyle{remark}

\usepackage[textsize=tiny]{todonotes}

\icmltitlerunning{An Unsupervised Learning Approach for Quasar Continuum Prediction}

\begin{document}

\twocolumn[
\icmltitle{An Unsupervised Learning Approach for Quasar Continuum Prediction}



\icmlsetsymbol{equal}{*}

\begin{icmlauthorlist}
\icmlauthor{Zechang Sun}{Tsinghua University}
\icmlauthor{Yuan-sen Ting}{ANU1,ANU2}
\icmlauthor{Zheng Cai}{Tsinghua University}
\end{icmlauthorlist}

\icmlaffiliation{Tsinghua University}{Department of Astronomy, Tsinghua University, Beijing, China}
\icmlaffiliation{ANU1}{Research School of Astronomy \& Astrophysics, Australian National University, Canberra, Australia}
\icmlaffiliation{ANU2}{School of Computing, Australian National University, Acton, ACT 2601, Australia}
\icmlcorrespondingauthor{Zechang Sun}{sunzc18@mails.tsinghua.edu.cn}

\icmlkeywords{Machine Learning, ICML}

\vskip 0.3in
]



\printAffiliationsAndNotice{}  

\begin{abstract}
Modeling quasar spectra is a fundamental task in astrophysics as quasars are the tell-tale sign of cosmic evolution. We introduce a novel unsupervised learning algorithm, Quasar Factor Analysis (QFA), for recovering the intrinsic quasar continua from noisy quasar spectra. QFA assumes that the Ly$\alpha$ forest can be approximated as a Gaussian process, and the continuum can be well described as a latent factor model. We show that QFA can learn, through unsupervised learning and directly from the quasar spectra, the quasar continua and Ly$\alpha$ forest simultaneously. Compared to previous methods, QFA achieves state-of-the-art performance for quasar continuum prediction robustly but without the need for predefined training continua. In addition, the generative and probabilistic nature of QFA paves the way to understanding the evolution of black holes as well as performing out-of-distribution detection and other Bayesian downstream inferences.
\end{abstract}

\section{Introduction}\label{submission}
Powered by the accretion of matter into the supermassive black holes in the galactic nuclei, luminous quasars have played a crucial role in astronomy as the lighthouses in the distant universe since their first discovery in the 1960s. Over the last 30 years, along with the unprecedented breakthroughs in quasar spectral observations \citep[e.g., SDSS Survey,][]{SDSS2020}, absorption systems in quasar spectra have been widely used to probe the state and matter distribution of intergalactic medium (IGM). The intervening gas between the quasar and us creates characteristic absorption superimposed on the quasar spectra, known as the Ly$\alpha$ forest. However, to extract information from the Ly$\alpha$ forest, one would first need to recover the intrinsic quasar continua. Any imperfection in continuum reconstruction can incur non-negligible uncertainties for any other downstream tasks, from inferring the cosmological parameters to understanding the physics of the IGM and constraining the astrophysics of reionization \citep{LYAPDF2011, 1Dpowerspectra2013, 1Dpowerspectra2019, Reionzation2020, compare2021}.

Due to its central role, different methods have been proposed to determine the intrinsic quasar continua, such as power-law explorations \citep{powerlaw2006}, principal component analysis \citep[PCA,][]{PCA2005,PCA2012,PCA2018} and deep learning techniques including multi-layer perceptron \citep[MLP,][]{QSmooth2020, iQNet2021} and normalizing flow \citep{normalizingFlow2020}. However, most methods proposed thus far belong to the nature of supervised learning. These methods are based on a training set of quasar spectra with pre-determined continua. As there is no way to know the ground truth continua, such continuum ``labels" are usually determined in an ad-hoc way through hand-fitting spectra with a high signal-to-noise ratio (SNR). Consequently, even though SDSS has observed more than 750,000 quasar spectra to date, typically only $0.01\%-1\%$ spectra with high SNR were used to construct the training set in previous works.

As the high SNR quasar spectra are often a biased subset of the entire quasar population, even with adequately supervised training, previous methods often struggle to generalize to all observed quasar spectra. These limitations invariably call for an unsupervised learning method, directly learning the distribution of the quasar spectra. However, unsupervised learning does not a priori lead to continuum; as quasar spectra are the composite of the quasar continuum, the intervening Ly$\alpha$ forest, and noise, these components are entirely degenerate. Thus, breaking this degeneracy requires us to harness the power of machine ``learning" while including sufficient physical prior for the ``modeling."

Here we propose Quasar Factor Analysis (QFA), an unsupervised algorithm that resolves these challenges that bottleneck the field: QFA can learn directly from millions of quasar spectra without any training continuum. It provides a fully probabilistic posterior of the continuum given the quasar spectrum. Furthermore, QFA provides physically meaningful spectrum embedding and enables us to perform out-of-distribution detection from noisy data.

\begin{figure*}[t]
\vskip 0.2in
\begin{center}
\centerline{\includegraphics[width=1.8\columnwidth]{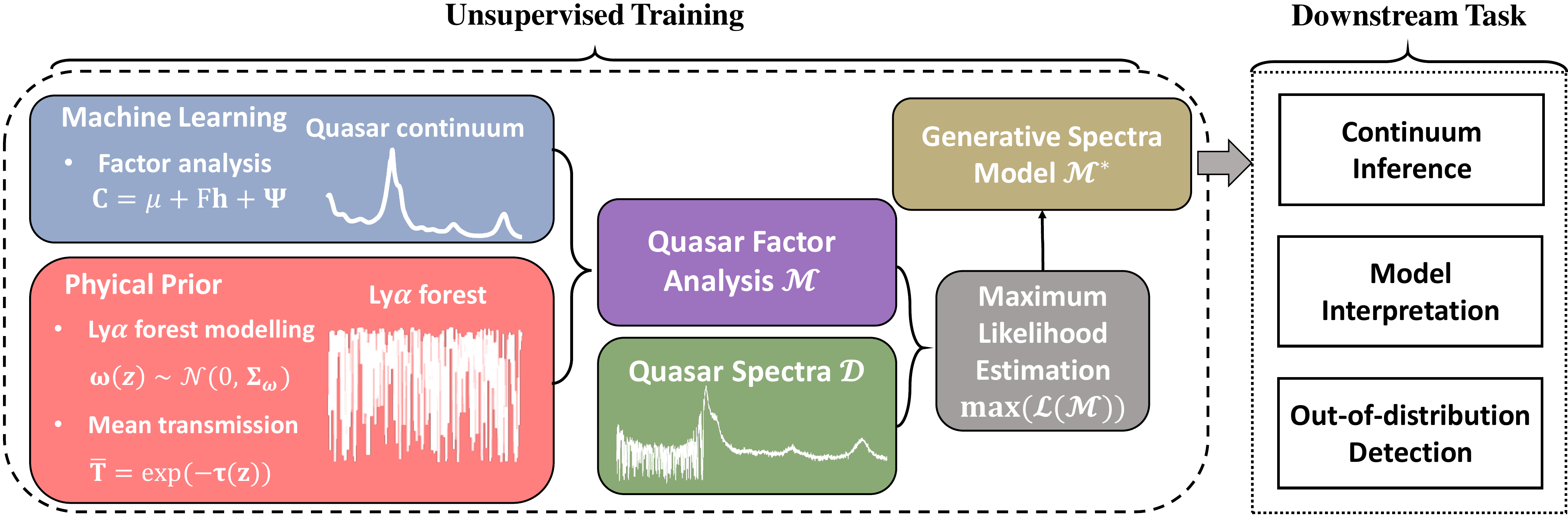}}
\caption{Model architecture. Quasar spectra consist of the intrinsic quasar continua and the Ly$\alpha$ forest. We show that by assuming a Gaussian process model on the Ly$\alpha$ forest and a linear latent model for the quasar continua, it is possible to model the distribution of the quasar spectra accurately through unsupervised learning, leading to a probabilistic way of performing other downstream tasks.}
\label{fig:pipeline}
\end{center}
\vskip -0.3in
\end{figure*}

\section{Method}

{\bf Statistical quasar model:} Let $\rm S$ be a quasar spectrum. We model the quasar spectrum as the composite of the continuum, $\rm C$, and the Ly$\alpha$ forest,
\begin{equation}
\rm S = C \circ \exp(-\tau(z)) + \omega(z) + \epsilon,
\end{equation}
where $\rm z$ is the redshift of the absorption systems, $\epsilon \sim \mathcal{N}(0,\Sigma_\epsilon)$ is the noise and,  ``$\circ$" denotes element-wise product. In this model, we specifically separate the continuum from the absorption features in the modeling. In this way, through unsupervised learning, we could learn the two components simultaneously, breaking the degeneracy between continuum and absorption.

The absorption is characterized by two terms, $\rm \exp(-\tau(z))$ and $\rm \omega(z)\sim \mathcal{N}(0,\Sigma_{\omega}(z))$. The former is commonly known as the mean transmission field and is fixed using the measurement from \citet{OpD2013}. On the other hand, $\rm \omega(z)$ is a learnable system that approximates the unaccounted random fluctuations caused by the absorbers. We further assume $\rm \Sigma_{\omega}$ is a diagonal matrix and
\begin{equation}
    \rm \text{diag}(\Sigma_{\omega}) = \omega_0\circ (1-\exp(-\tau_0 (1+z)^\beta )+c_0)^2,
\end{equation}
\noindent
where $\rm \omega_0$, $\rm \tau_0$, $\rm \beta$, $\rm c_0$ are free parameters which we will optimize for.

Previous works largely inspired this particular Ly$\alpha$ model \cite{GP2020, PICCA2020} which show that modeling the absorption field as a Gaussian process with the absorber redshift dependency is a robust description of the quasar absorption. Although not explicitly shown, we assume that absorption terms only apply to wavelength bluer than the quasar Ly$\alpha$ emission as the Universe expands and redshifts the quasar continuum. The Ly$\alpha$ forest only imprints on the blue part of the quasar spectrum.

As for the continuum $\rm C$, previous PCA based methods \cite{PCA2005, PCA2012, PCA2018}, have demonstrated that the quasar continua can be robustly captured with a linear latent model. Guided by these successes, we describe the quasar continua with a latent factor model as follows:
\begin{equation}
    \rm C = \mu + Fh + \Psi,
\end{equation}
generalizing PCA to a probabilistic model while going beyond the orthonormal basis as assumed in the PCA formalism. $\rm \mu$ is the mean vector. $\rm \Psi\sim\mathcal{N}(0, \Sigma_{\Psi})$ represents the ``stochastic" term capturing variances that are not modeled by the latent model, and  $\rm F\in\mathbb{R}^{N_{s}\times N_h}$ is the factor loading matrix. Here $\rm N_s$ is the dimension of spectra, $\rm N_h$ is the dimension of hidden variable and $\rm N_h \ll N_s$. All of these are learnable parameters that will be optimized through maximum likelihood estimation. Moreover, $\rm h\sim\mathcal{N}(0, I)$ is the factor or hidden variable with a much lower dimension (here we chose it to be eight), for which we will infer its posterior.

Finally, let $\rm A$ be the diagonal matrix of $\rm \exp(-\tau(z))$, it follows that the distribution of the quasar spectrum $\rm S$ can be approximated as:
\begin{equation}
    \rm S\sim\mathcal{N}( A\mu, A(FF^T+\Sigma_{\Psi})A^T+\Sigma_{\omega}+\Sigma_{\epsilon}).
    \label{eq:dis}
\end{equation}\par
\begin{figure*}[t]
\vskip 0.2in
\begin{center}
\centerline{\includegraphics[width=1.9\columnwidth]{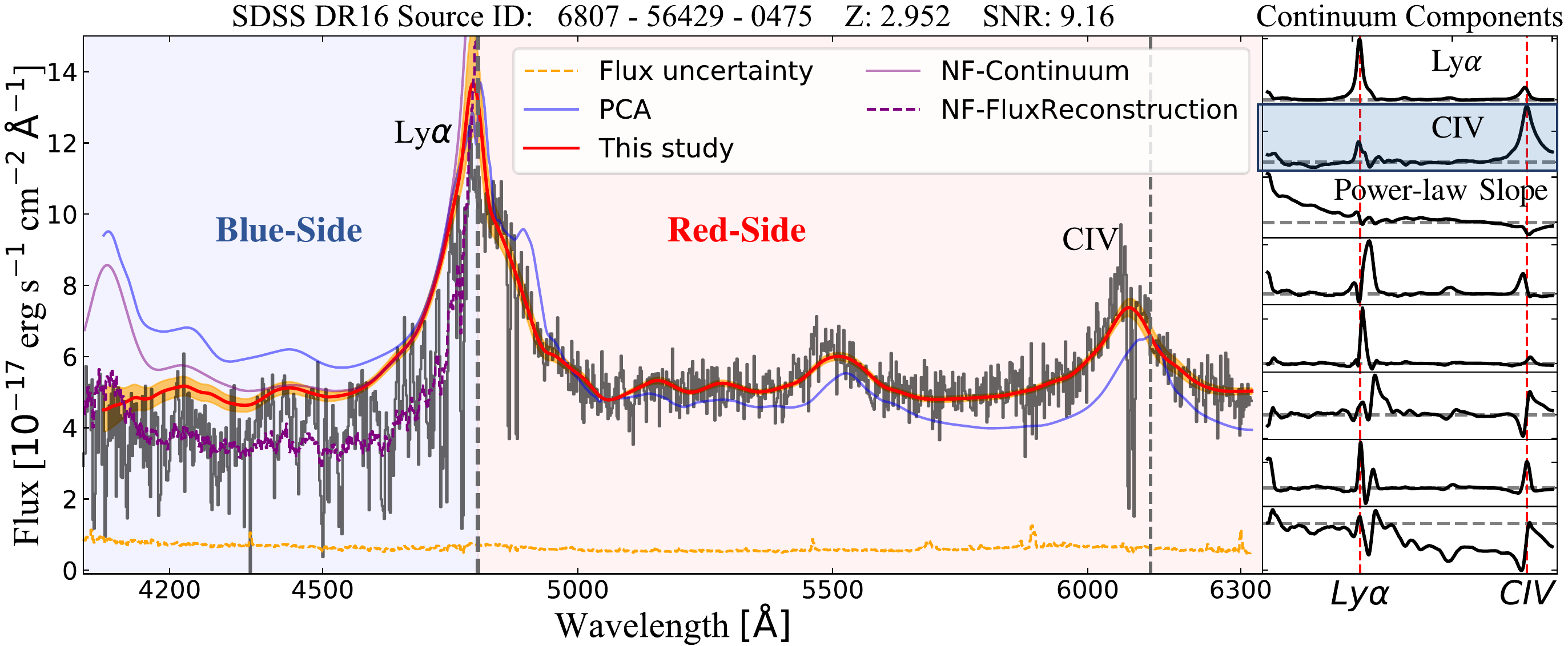}}
\caption{{\bf Left:} QFA provides a probabilistic output and describes the continuum more sensibly than PCA. Shown in black is an SDSS quasar spectrum. The Orange region denotes the $95\%$ confidence interval for QFA posterior prediction. The PCA model trained on high SNR training spectra fails to generalize to lower SNR samples, whereas QFA recovers the continuum accurately. In addition, the unsupervised deep learning method (normalizing flows or NF) fails as it suffers from mode collapse, and the reconstructed flux converges to mean spectrum. Such failure leads to problematic continuum inference with NF. {\bf Right:} The continuum components learned by QFA are physically sensible, including one probing the Ly$\alpha$ emission and the other a power-law slope reminiscent of \citet{powerlaw2006}. The CIV component shaded in blue shows the Baldwin effect (see text).}
\label{fig:example}
\end{center}
\vskip -0.3in
\end{figure*}
{\bf Maximum likelihood optimization:} Given our statistical model of the quasar spectra, it then follows that for any observed set of $\rm N$ training quasar spectra with their measurement uncertainties $\mathcal{D} = \rm \{S^{(i)}, \Sigma^{(i)}_\epsilon, z^{(i)}\}$, we can optimize the model parameters $\rm \mathcal{M} = \{\mu, F, \Sigma_{\Psi}, \omega_0, \tau_0,
\beta, c_0\}$ simultaneously through maximizing the likelihood. In particular, the log-likelihood of all the observations can be written as
\begin{equation}
    \rm \mathcal{L}(\mathcal{M}) = \frac{1}{N}\sum_{i=1}^{N}\log Pr( S^{(i)}|\Sigma_{\epsilon}^{(i)}, z^{(i)},\mathcal{M}),
    \label{eq:loss}
\end{equation}
We note that training such a non-conventional latent factor model with multiple massive moving parts cannot be solved with the classical EM method \cite{FA2012} but is made possible with the modern-day deep learning framework. We develop our codes in Pytorch. The codes are available on Github. The models are optimized via \texttt{Adam} algorithm \cite{Adam2014} to search for a local minimum $\mathcal{M}^*$.

{\bf Regularization and tricks:} Due to the large number of model parameters ($\sim 2\times10^4$) and the strong degeneracy between continuum, absorption, and noise, carefully designed model regularization and optimization procedures are needed to ensure that QFA converges. First, as the variations between different continua are relatively small compared to the mean continuum, we expect the learned continuum components to have small deviations from zero. This inspires us to assign a $\rm L2$ regularization to model parameters. We further avoid model singularity by setting minimum values of $\Sigma_{\Psi}$ and $\Sigma_{\omega}$
as well as early stopping when the average negative log-likelihood is smaller than zero. Finally, to enforce that the predicted continua are smooth radiation profiles, we smooth each continuum component with a running window every 20 epochs during training.

{\bf Continuum inference:} Once the model parameters are optimized, recall that the continuum is modeled as $\rm \mu + Fh$. Therefore, to drive the probabilistic output of the continuum, it suffices to derive the posterior distribution of the hidden variable $\rm h$ given the observed quasar spectrum $\rm S$, or mathematically,
\begin{equation}
       \rm {\rm P}(h|\rm z, S, \Sigma_{\epsilon}, \mathcal{M}^*) \propto {\rm P}(S|\rm z, h, \Sigma_{\epsilon}, \mathcal{M}^*) {\rm P}(h).
        \label{eq:posh}
\end{equation}
Here we assume a standard normal distribution for the prior of $\rm h$, or $\rm h \sim\mathcal{N}(0, I)$.

\section{Data}
We evaluate model performance both on SDSS quasar spectra and mock quasar spectra. For the SDSS data set, we select $90,606$ quasar spectra with SNR greater than 2 and redshift from 2 to 3.5 from the SDSS data release 16 quasar catalogue \cite{SDSS2020}. To numerically evaluate model performance, we also generate mock quasar spectra with which the ground truth continua are known. More specifically, we measure continua of the SDSS data set with PCA components given by \citet{Component2011}, learn the distribution of the PCA coefficients via a Gaussian Mixture Model and sample new continua from the distribution of the coefficients. As for the mock absorption field, we adopt the well-calibrated SDSS data release 11 quasar-Ly$\alpha$ mock data set \citep{SDSSMOCK2015}. We further add Gaussian random noise to mock spectra to mimic observations. Besides, to demonstrate that our model can generalize to other quasar spectra outside the high SNR data with which the PCA model was trained, we also assume $\lesssim 10\%$ random linear perturbations to the mock continua. Our final mock data set consists of $\sim$ 150,000 quasar spectra.

\section{Results}
{\bf Comparison to classical methods:} We first compare QFA's performance on the mock data set with the more widely adopted PCA algorithm \citep[hereafter PCA,][]{Component2011}. By fitting all $\sim$150,000 mock spectra, we found that when fitting mock spectra are generated from the PCA basis without perturbation, both QFA and PCA reach a median $\sim 2\%$ relative error  (integrated over all pixels). Our result demonstrates that even without any training continuum, QFA can perform on par with PCA, recovering continua from noisy data based entirely on unsupervised learning.

More importantly, as we perturb the mock continua, PCA fails to generalize and incurs a relative error of $\sim 4\%$, while QFA maintains the same level of performance, at a $\sim 2\%$ relative error on the blue side. Furthermore, on the red side, QFA can reach a relative error of $\lesssim 1\%$, but PCA often incurs a non-negligible $\sim 3-4\%$ relative error.

Fig.~\ref{fig:example} further demonstrates that QFA also performs well on actual SDSS quasar spectra. As shown in Fig.~\ref{fig:example}, PCA fails to generalize to some low SNR quasar spectra; the biased training based on high SNR leads to a nonphysical continuum. However, QFA performs well and achieves a more sensible continuum, consistent with the mock test.

{\bf Model interpretability:} As QFA assumes a linear latent model for the continuum, it decomposes quasar continua into the key contributing components. Applying QFA to the actual SDSS DR 16 quasar spectra appears to lead to physically sensible components. As shown in Fig.~\ref{fig:example}, these individual components include Ly$\alpha$ emission, CIV emission, and the power-law slope components. Our model learns these components directly from the quasar spectra without training continua, thus not biasing us to only the high SNR and hence, low-redshift quasar data. We found that, within the redshift range covered by SDSS DR16 ($\rm z=2-3.5$), the individual physical components of the quasar continua have no detectable evolution with redshift, demonstrating that the quasar population has not evolved much during this period. Additionally, QFA reveals also a well-documented correlation known as the Baldwin effect \cite{Baldwin1977}. The Baldwin effect posits that the component corresponding to CIV emission strength negatively correlates with the quasar monochromatic luminosity, which QFA bears out.

\begin{figure}
\vskip 0.1in
\begin{center}
\centerline{\includegraphics[width=\columnwidth]{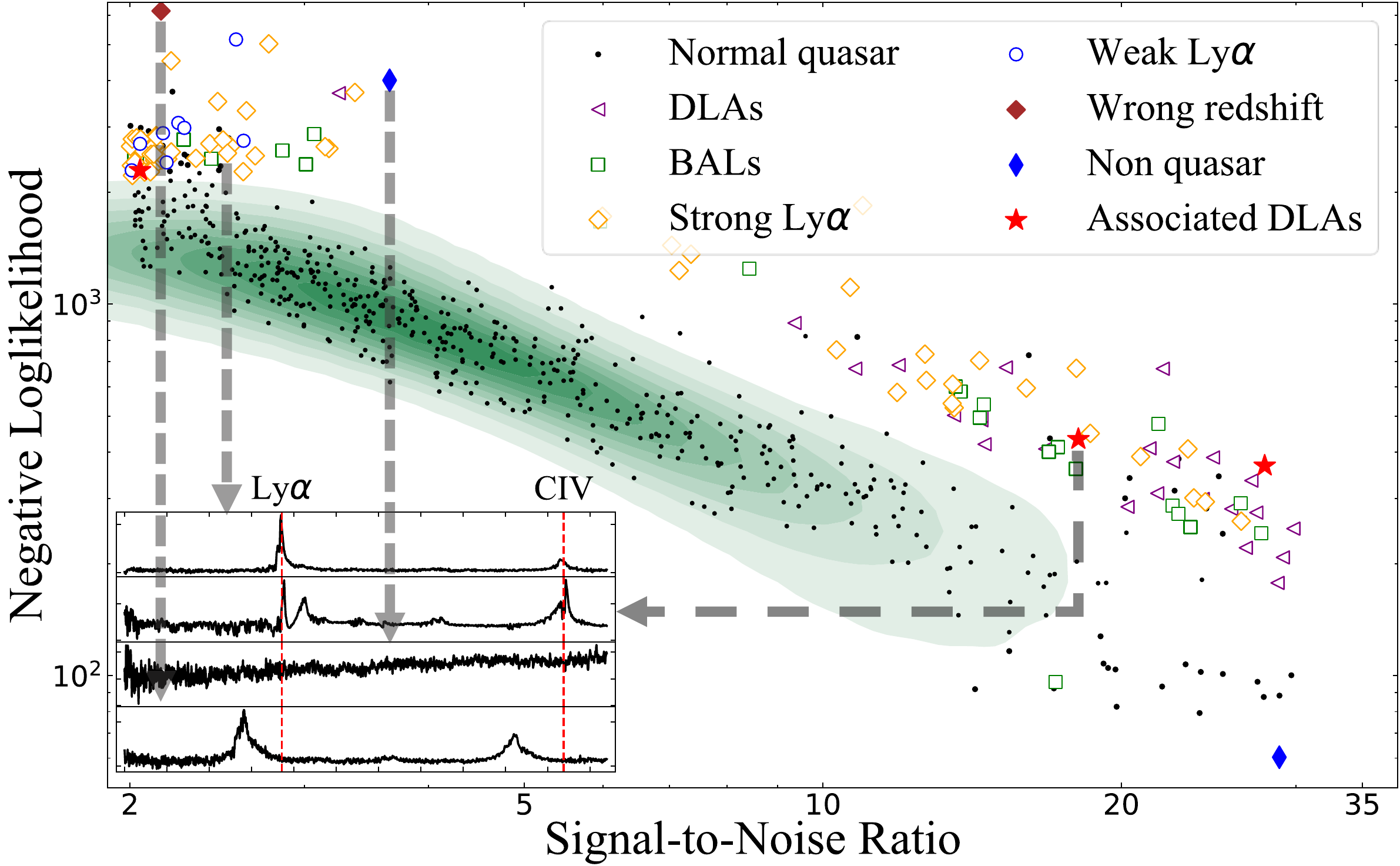}}
\caption{The generative and probabilistic nature of QFA allows it to perform outlier detection effectively. The background contour shows the likelihood of all the SDSS DR16 spectra. Color-coded at the low-likelihood peripheric regions shows some of the outlier objects of interest, classified by their respective classes. The bottom left corner exemplifies the spectra of some of these outliers.}
\label{fig:outlier}
\end{center}
\vskip -0.2in
\end{figure}

{\bf Outlier detection:} Finally, QFA's probabilistic nature enables also out-of-distribution detection. Specifically, for any observed spectrum, the likelihood can be evaluated with Equation~\ref{eq:dis}. We applied our method to the SDSS DR16 data and performed outlier detection following \citet{PYOD2019}. As shown in Fig.~\ref{fig:outlier}, QFA unearthed multiple outliers, a few of them were previously unknown, including (a) undetected damped Ly$\alpha$ absorbers (DLAs); (b) associated damped Ly$\alpha$ absorbers (associated DLAs); (c) broad absorption lines (BALs); (d) Type II quasar feature -- strong Ly$\alpha$ emission but weak continuum; (e) wrong redshift estimation; (f) misclassified non-quasar spectra. These outliers will be explored in detail in our forthcoming paper.

\section{Discussion and Conclusion}
{\bf Deep learning?} As our method is rooted in describing the distribution of quasar spectra, a natural question arises: Can more sophisticated deep learning generative models better describe the quasar spectra distribution? Applying deep learning was the first attempt in this study. In particular, we assumed two separate normalizing flows \cite{VAENF2015} to capture the continua and the absorption features. We pre-trained each normalizing flow to learn their respective domain and then perform variational inference on real spectra, with the hope that the two normalizing flows will domain-adapt to any synthetic gap in their respective domain. However, the lack of physical prior often leads to the reconstruction collapsing to the mean spectrum, as shown in Fig.~\ref{fig:example}. The inability to break the degeneracy subsequently leads to poor continuum reconstruction. Thus, the lack of rigidity on the physical prior and the high model complexity render unsupervised deep learning infeasible in this particular context.

{\bf In Summary}, we demonstrate that we could recover the quasar continua robustly without any training continua. QFA harnesses the entire data set with heteroscedastic noise through unsupervised learning and can reach state-of-the-art performance for quasar continuum recovery. With more quasar spectra being collected from ongoing spectroscopic surveys such as DESI and SDSS-V, QFA might prove critical. It does not rely on continuum labels and can automatically adapt to latent variations in the data set. But perhaps more interesting, in a time where deep learning has reigned supreme in many areas of astronomy, our work sheds light that classical machine ``learning" with physical ``modeling" can remain powerful, harnessing the best of both worlds.

\nocite{langley00}

\bibliography{example_paper}

\begin{thebibliography}{23}
\providecommand{\natexlab}[1]{#1}
\providecommand{\url}[1]{\texttt{#1}}
\expandafter\ifx\csname urlstyle\endcsname\relax
  \providecommand{\doi}[1]{doi: #1}\else
  \providecommand{\doi}{doi: \begingroup \urlstyle{rm}\Url}\fi

\bibitem[{Baldwin}(1977)]{Baldwin1977}
{Baldwin}, J.~A.
\newblock {Luminosity Indicators in the Spectra of Quasi-Stellar Objects}.
\newblock \emph{The Astrophysical Journal}, 214:\penalty0 679--684, June 1977.
\newblock \doi{10.1086/155294}.

\bibitem[Barber(2012)]{FA2012}
Barber, D.
\newblock \emph{Bayesian Reasoning and Machine Learning}.
\newblock Cambridge University Press, 2012.
\newblock \doi{10.1017/CBO9780511804779}.

\bibitem[Bautista et~al.(2015)Bautista, Bailey, Font-Ribera, Pieri, Busca,
  Miralda-Escud{\'{e}}, Palanque-Delabrouille, Rich, Dawson, Feng, Ge, Gontcho,
  Ho, Goff, Noterdaeme, P{\^{a}}ris, Rossi, and Schlegel]{SDSSMOCK2015}
Bautista, J.~E., Bailey, S., Font-Ribera, A., Pieri, M.~M., Busca, N.~G.,
  Miralda-Escud{\'{e}}, J., Palanque-Delabrouille, N., Rich, J., Dawson, K.,
  Feng, Y., Ge, J., Gontcho, S. G.~A., Ho, S., Goff, J. M.~L., Noterdaeme, P.,
  P{\^{a}}ris, I., Rossi, G., and Schlegel, D.
\newblock Mock quasar-lyman-$\alpha$ forest data-sets for the {SDSS}-{III}
  baryon oscillation spectroscopic survey.
\newblock \emph{Journal of Cosmology and Astroparticle Physics}, 2015\penalty0
  (05):\penalty0 060--060, may 2015.
\newblock \doi{10.1088/1475-7516/2015/05/060}.
\newblock URL \url{https://doi.org/10.1088/1475-7516/2015/05/060}.

\bibitem[{Becker} et~al.(2013){Becker}, {Hewett}, {Worseck}, and
  {Prochaska}]{OpD2013}
{Becker}, G.~D., {Hewett}, P.~C., {Worseck}, G., and {Prochaska}, J.~X.
\newblock {A refined measurement of the mean transmitted flux in the
  Ly${\ensuremath{\alpha}}$ forest over $2 < z < 5$ using composite quasar
  spectra}.
\newblock \emph{Monthly Notices of the Royal Astronomical Society},
  430\penalty0 (3):\penalty0 2067--2081, April 2013.
\newblock \doi{10.1093/mnras/stt031}.

\bibitem[Bosman et~al.(2021)Bosman, {\v{D} }urov{\v{c}}{\'{\i}}kov{\'{a}},
  Davies, and Eilers]{compare2021}
Bosman, S. E.~I., {\v{D} }urov{\v{c}}{\'{\i}}kov{\'{a}}, D., Davies, F.~B., and
  Eilers, A.-C.
\newblock A comparison of quasar emission reconstruction techniques for $\rm z
  \geq 5.0$ ly$\alpha$ and ly$\beta$ transmission.
\newblock \emph{Monthly Notices of the Royal Astronomical Society},
  503\penalty0 (2):\penalty0 2077--2096, feb 2021.
\newblock \doi{10.1093/mnras/stab572}.
\newblock URL \url{https://doi.org/10.1093%2Fmnras%2Fstab572}.

\bibitem[Chabanier et~al.(2019)Chabanier, Palanque-Delabrouille, Y{\`e}che,
  Le~Goff, Armengaud, Bautista, Blomqvist, Dawson, Etourneau, Font-Ribera,
  et~al.]{1Dpowerspectra2019}
Chabanier, S., Palanque-Delabrouille, N., Y{\`e}che, C., Le~Goff, J.-M.,
  Armengaud, E., Bautista, J., Blomqvist, M., Dawson, K., Etourneau, T.,
  Font-Ribera, A., et~al.
\newblock The one-dimensional power spectrum from the sdss dr14 ly$\alpha$
  forests.
\newblock \emph{Journal of Cosmology and Astroparticle Physics}, 2019\penalty0
  (07):\penalty0 017, 2019.

\bibitem[Davies et~al.(2018)Davies, Hennawi, Ba{\~{n} }ados, Luki{\'{c}},
  Decarli, Fan, Farina, Mazzucchelli, Rix, Venemans, Walter, Wang, and
  Yang]{PCA2018}
Davies, F.~B., Hennawi, J.~F., Ba{\~{n} }ados, E., Luki{\'{c}}, Z., Decarli,
  R., Fan, X., Farina, E.~P., Mazzucchelli, C., Rix, H.-W., Venemans, B.~P.,
  Walter, F., Wang, F., and Yang, J.
\newblock Quantitative constraints on the reionization history from the {IGM}
  damping wing signature in two quasars at z > 7.
\newblock \emph{The Astrophysical Journal}, 864\penalty0 (2):\penalty0 142, sep
  2018.
\newblock \doi{10.3847/1538-4357/aad6dc}.
\newblock URL \url{https://doi.org/10.3847%2F1538-4357%2Faad6dc}.

\bibitem[du~Mas~des Bourboux et~al.(2020)du~Mas~des Bourboux, Rich,
  Font-Ribera, de~Sainte~Agathe, Farr, Etourneau, Le~Goff, Cuceu, Balland,
  Bautista, and et~al.]{PICCA2020}
du~Mas~des Bourboux, H., Rich, J., Font-Ribera, A., de~Sainte~Agathe, V., Farr,
  J., Etourneau, T., Le~Goff, J.-M., Cuceu, A., Balland, C., Bautista, J.~E.,
  and et~al.
\newblock The completed sdss-iv extended baryon oscillation spectroscopic
  survey: Baryon acoustic oscillations with ly$\alpha$ forests.
\newblock \emph{The Astrophysical Journal}, 901\penalty0 (2):\penalty0 153, Oct
  2020.
\newblock ISSN 1538-4357.
\newblock \doi{10.3847/1538-4357/abb085}.
\newblock URL \url{http://dx.doi.org/10.3847/1538-4357/abb085}.

\bibitem[{Fan} et~al.(2006){Fan}, {Strauss}, {Becker}, {White}, {Gunn},
  {Knapp}, {Richards}, {Schneider}, {Brinkmann}, and {Fukugita}]{powerlaw2006}
{Fan}, X., {Strauss}, M.~A., {Becker}, R.~H., {White}, R.~L., {Gunn}, J.~E.,
  {Knapp}, G.~R., {Richards}, G.~T., {Schneider}, D.~P., {Brinkmann}, J., and
  {Fukugita}, M.
\newblock {Constraining the Evolution of the Ionizing Background and the Epoch
  of Reionization with z\raisebox{-0.5ex}\textasciitilde6 Quasars. II. A Sample
  of 19 Quasars}.
\newblock \emph{\aj}, 132\penalty0 (1):\penalty0 117--136, July 2006.
\newblock \doi{10.1086/504836}.

\bibitem[Ho et~al.(2020)Ho, Bird, and Garnett]{GP2020}
Ho, M.-F., Bird, S., and Garnett, R.
\newblock Detecting multiple dlas per spectrum in sdss dr12 with gaussian
  processes.
\newblock \emph{Monthly Notices of the Royal Astronomical Society},
  496\penalty0 (4):\penalty0 5436–5454, Jun 2020.
\newblock ISSN 1365-2966.
\newblock \doi{10.1093/mnras/staa1806}.
\newblock URL \url{http://dx.doi.org/10.1093/mnras/staa1806}.

\bibitem[Kingma \& Ba(2014)Kingma and Ba]{Adam2014}
Kingma, D.~P. and Ba, J.
\newblock Adam: A method for stochastic optimization.
\newblock \emph{arXiv preprint arXiv:1412.6980}, 2014.

\bibitem[Lee \& Spergel(2011)Lee and Spergel]{LYAPDF2011}
Lee, K.-G. and Spergel, D.~N.
\newblock Threshold probability functions and thermal inhomogeneities in the
  ly$\alpha$ forest.
\newblock \emph{The Astrophysical Journal}, 734\penalty0 (1):\penalty0 21, may
  2011.
\newblock \doi{10.1088/0004-637x/734/1/21}.
\newblock URL \url{https://doi.org/10.1088%2F0004-637x%2F734%2F1%2F21}.

\bibitem[Lee et~al.(2012)Lee, Suzuki, and Spergel]{PCA2012}
Lee, K.-G., Suzuki, N., and Spergel, D.~N.
\newblock Mean-flux-regulated principal component analysis continuum fitting of
  sloan digital sky survey ly$\alpha$ forest spectra.
\newblock \emph{The Astronomical Journal}, 143\penalty0 (2):\penalty0 51, Jan
  2012.
\newblock ISSN 1538-3881.
\newblock \doi{10.1088/0004-6256/143/2/51}.
\newblock URL \url{http://dx.doi.org/10.1088/0004-6256/143/2/51}.

\bibitem[Liu \& Bordoloi(2021)Liu and Bordoloi]{iQNet2021}
Liu, B. and Bordoloi, R.
\newblock A deep learning approach to quasar continuum prediction.
\newblock \emph{Monthly Notices of the Royal Astronomical Society},
  502\penalty0 (3):\penalty0 3510–3532, Jan 2021.
\newblock ISSN 1365-2966.
\newblock \doi{10.1093/mnras/stab177}.
\newblock URL \url{http://dx.doi.org/10.1093/mnras/stab177}.

\bibitem[Lyke et~al.(2020)Lyke, Higley, McLane, Schurhammer, Myers, Ross,
  Dawson, Chabanier, Martini, Des~Bourboux, et~al.]{SDSS2020}
Lyke, B.~W., Higley, A.~N., McLane, J., Schurhammer, D.~P., Myers, A.~D., Ross,
  A.~J., Dawson, K., Chabanier, S., Martini, P., Des~Bourboux, H. D.~M., et~al.
\newblock The sloan digital sky survey quasar catalog: Sixteenth data release.
\newblock \emph{The Astrophysical Journal Supplement Series}, 250\penalty0
  (1):\penalty0 8, 2020.

\bibitem[Montero-Camacho \& Mao(2020)Montero-Camacho and Mao]{Reionzation2020}
Montero-Camacho, P. and Mao, Y.
\newblock Ly{\hspace{0.167em} }$\alpha$ forest power spectrum as an emerging
  window into the epoch of reionization and cosmic dawn.
\newblock \emph{Monthly Notices of the Royal Astronomical Society},
  499\penalty0 (2):\penalty0 1640--1651, sep 2020.
\newblock \doi{10.1093/mnras/staa2918}.
\newblock URL \url{https://doi.org/10.1093\%2Fmnras\%2Fstaa2918}.

\bibitem[Palanque-Delabrouille et~al.(2013)Palanque-Delabrouille, Y{\`e}che,
  Borde, Le~Goff, Rossi, Viel, Aubourg, Bailey, Bautista, Blomqvist,
  et~al.]{1Dpowerspectra2013}
Palanque-Delabrouille, N., Y{\`e}che, C., Borde, A., Le~Goff, J.-M., Rossi, G.,
  Viel, M., Aubourg, {\'E}., Bailey, S., Bautista, J., Blomqvist, M., et~al.
\newblock The one-dimensional ly$\alpha$ forest power spectrum from boss.
\newblock \emph{Astronomy \& Astrophysics}, 559:\penalty0 A85, 2013.

\bibitem[Pâris et~al.(2011)Pâris, Petitjean, Rollinde, Aubourg, Busca,
  Charlassier, Delubac, Hamilton, Le~Goff, Palanque-Delabrouille, and
  et~al.]{Component2011}
Pâris, I., Petitjean, P., Rollinde, E., Aubourg, E., Busca, N., Charlassier,
  R., Delubac, T., Hamilton, J.-C., Le~Goff, J.-M., Palanque-Delabrouille, N.,
  and et~al.
\newblock A principal component analysis of quasar uv spectra atz ~  3.
\newblock \emph{Astronomy \& Astrophysics}, 530:\penalty0 A50, May 2011.
\newblock ISSN 1432-0746.
\newblock \doi{10.1051/0004-6361/201016233}.
\newblock URL \url{http://dx.doi.org/10.1051/0004-6361/201016233}.

\bibitem[Reiman et~al.(2020)Reiman, Tamanas, Prochaska, and
  Ďurovčíková]{normalizingFlow2020}
Reiman, D.~M., Tamanas, J., Prochaska, J.~X., and Ďurovčíková, D.
\newblock Fully probabilistic quasar continua predictions near lyman-$\alpha$
  with conditional neural spline flows, 2020.
\newblock URL \url{https://arxiv.org/abs/2006.00615}.

\bibitem[Rezende \& Mohamed(2015)Rezende and Mohamed]{VAENF2015}
Rezende, D.~J. and Mohamed, S.
\newblock Variational inference with normalizing flows, 2015.
\newblock URL \url{https://arxiv.org/abs/1505.05770}.

\bibitem[Suzuki et~al.(2005)Suzuki, Tytler, Kirkman, O'Meara, and
  Lubin]{PCA2005}
Suzuki, N., Tytler, D., Kirkman, D., O'Meara, J.~M., and Lubin, D.
\newblock Predicting {QSO} continua in the ly$\alpha$ forest.
\newblock \emph{The Astrophysical Journal}, 618\penalty0 (2):\penalty0
  592--600, jan 2005.
\newblock \doi{10.1086/426062}.
\newblock URL \url{https://doi.org/10.1086/426062}.

\bibitem[Zhao et~al.(2019)Zhao, Nasrullah, and Li]{PYOD2019}
Zhao, Y., Nasrullah, Z., and Li, Z.
\newblock Pyod: A python toolbox for scalable outlier detection.
\newblock \emph{Journal of Machine Learning Research}, 20\penalty0
  (96):\penalty0 1--7, 2019.
\newblock URL \url{http://jmlr.org/papers/v20/19-011.html}.

\bibitem[Ďurovčíková et~al.(2020)Ďurovčíková, Katz, Bosman, Davies,
  Devriendt, and Slyz]{QSmooth2020}
Ďurovčíková, D., Katz, H., Bosman, S. E.~I., Davies, F.~B., Devriendt, J.,
  and Slyz, A.
\newblock Reionization history constraints from neural network based
  predictions of high-redshift quasar continua.
\newblock \emph{Monthly Notices of the Royal Astronomical Society},
  493\penalty0 (3):\penalty0 4256–4275, Feb 2020.
\newblock ISSN 1365-2966.
\newblock \doi{10.1093/mnras/staa505}.
\newblock URL \url{http://dx.doi.org/10.1093/mnras/staa505}.

\end{thebibliography}
\bibliographystyle{icml2022}
\end{document}